\documentclass[10pt,a4paper,onecolumn]{article}
\usepackage{marginnote}
\usepackage{graphicx}
\usepackage{xcolor}
\usepackage{authblk,etoolbox}
\usepackage{titlesec}
\usepackage{calc}
\usepackage{tikz}
\usepackage{hyperref}
\hypersetup{colorlinks,breaklinks=true,
            urlcolor=[rgb]{0.0, 0.5, 1.0},
            linkcolor=[rgb]{0.0, 0.5, 1.0}}
\usepackage{caption}
\usepackage{tcolorbox}
\usepackage{amssymb,amsmath}
\usepackage{ifxetex,ifluatex}
\usepackage{seqsplit}
\usepackage{xstring}

\usepackage{float}
\let\origfigure\figure
\let\endorigfigure\endfigure
\renewenvironment{figure}[1][2] {
    \expandafter\origfigure\expandafter[H]
} {
    \endorigfigure
}

\usepackage{fixltx2e} 
\usepackage[
  backend=biber,
]{biblatex}
\bibliography{bib.bib}

\let\textttOrig=\texttt
\def\texttt#1{\expandafter\textttOrig{\seqsplit{#1}}}
\renewcommand{\seqinsert}{\ifmmode
  \allowbreak
  \else\penalty6000\hspace{0pt plus 0.02em}\fi}

\makeatletter
\let\href@Orig=\href
\def\href@Urllike#1#2{\href@Orig{#1}{\begingroup
    \def\Url@String{#2}\Url@FormatString
    \endgroup}}
\def\href@Notdoi#1#2{\def\tempa{#1}\def\tempb{#2}%
  \ifx\tempa\tempb\relax\href@Urllike{#1}{#2}\else
  \href@Orig{#1}{#2}\fi}
\def\href#1#2{%
  \IfBeginWith{#1}{https://doi.org}%
  {\href@Urllike{#1}{#2}}{\href@Notdoi{#1}{#2}}}
\makeatother

\newlength{\cslhangindent}
\newlength{\csllabelwidth}
\newenvironment{CSLReferences}[3] 
 {
  \setlength{\parindent}{0pt}
  \ifodd #1 \everypar{\setlength{\hangindent}{\cslhangindent}}\ignorespaces\fi
  \ifnum #2 > 0
  \setlength{\parskip}{#2\baselineskip}
  \fi
 }%
 {}
\usepackage{calc}

\usepackage[top=3.5cm, bottom=3cm, right=1.5cm, left=1.0cm,
            headheight=2.2cm, reversemp, includemp, marginparwidth=4.5cm]{geometry}

\titleformat{\section}
  {\normalfont\sffamily\Large\bfseries}
  {}{0pt}{}
\titleformat{\subsection}
  {\normalfont\sffamily\large\bfseries}
  {}{0pt}{}
\titleformat{\subsubsection}
  {\normalfont\sffamily\bfseries}
  {}{0pt}{}
\titleformat*{\paragraph}
  {\sffamily\normalsize}

\usepackage{fancyhdr}
\makeatletter
\let\ps@plain\ps@fancy
\fancyheadoffset[L]{4.5cm}
\fancyfootoffset[L]{4.5cm}

\definecolor{linky}{rgb}{0.0, 0.5, 1.0}

\newtcolorbox{repobox}
   {colback=red, colframe=red!75!black,
     boxrule=0.5pt, arc=2pt, left=6pt, right=6pt, top=3pt, bottom=3pt}

\newcommand{\ExternalLink}{%
   \tikz[x=1.2ex, y=1.2ex, baseline=-0.05ex]{%
       \begin{scope}[x=1ex, y=1ex]
           \clip (-0.1,-0.1)
               --++ (-0, 1.2)
               --++ (0.6, 0)
               --++ (0, -0.6)
               --++ (0.6, 0)
               --++ (0, -1);
           \path[draw,
               line width = 0.5,
               rounded corners=0.5]
               (0,0) rectangle (1,1);
       \end{scope}
       \path[draw, line width = 0.5] (0.5, 0.5)
           -- (1, 1);
       \path[draw, line width = 0.5] (0.6, 1)
           -- (1, 1) -- (1, 0.6);
       }
   }

\patchcmd{\@maketitle}{center}{flushleft}{}{}
\patchcmd{\@maketitle}{center}{flushleft}{}{}
\patchcmd{\@maketitle}{\LARGE}{\LARGE\sffamily}{}{}
\def\maketitle{{%
  
  \AB@maketitle}}
\makeatletter
\renewcommand\AB@affilsepx{ \protect\Affilfont}
\renewcommand\AB@affilnote[1]{{\bfseries #1}\hspace{3pt}}
\renewcommand{\affil}[2][]%
   {\newaffiltrue\let\AB@blk@and\AB@pand
      \if\relax#1\relax\def\AB@note{\AB@thenote}\else\def\AB@note{#1}%
        \setcounter{Maxaffil}{0}\fi
        \begingroup
        \let\href=\href@Orig
        \let\texttt=\textttOrig
        \let\protect\@unexpandable@protect
        \def\thanks{\protect\thanks}\def\footnote{\protect\footnote}%
        \@temptokena=\expandafter{\AB@authors}%
        {\def\\{\protect\\\protect\Affilfont}\xdef\AB@temp{#2}}%
         \xdef\AB@authors{\the\@temptokena\AB@las\AB@au@str
         \protect\\[\affilsep]\protect\Affilfont\AB@temp}%
         \gdef\AB@las{}\gdef\AB@au@str{}%
        {\def\\{, \ignorespaces}\xdef\AB@temp{#2}}%
        \@temptokena=\expandafter{\AB@affillist}%
        \xdef\AB@affillist{\the\@temptokena \AB@affilsep
          \AB@affilnote{\AB@note}\protect\Affilfont\AB@temp}%
      \endgroup
       \let\AB@affilsep\AB@affilsepx
}
\makeatother

\renewcommand\Affilfont{\sffamily\small\mdseries}
\ifnum 0\ifxetex 1\fi\ifluatex 1\fi=0 
  \usepackage[T1]{fontenc}
  \usepackage[utf8]{inputenc}

\else 
  \ifxetex
    \usepackage{mathspec}
    \usepackage{fontspec}

  \else
    \usepackage{fontspec}
  \fi
  \defaultfontfeatures{Ligatures=TeX,Scale=MatchLowercase}

\fi
\IfFileExists{upquote.sty}{\usepackage{upquote}}{}
\IfFileExists{microtype.sty}{%
\usepackage{microtype}
\UseMicrotypeSet[protrusion]{basicmath} 
}{}

\usepackage{hyperref}
\hypersetup{unicode=true,
            pdftitle={starry\_process: Interpretable Gaussian processes for stellar light curves},
            pdfborder={0 0 0},
            breaklinks=true}
\let\addcontentslineOrig=\addcontentsline
\def\addcontentsline#1#2#3{\bgroup
  \let\texttt=\textttOrig\addcontentslineOrig{#1}{#2}{#3}\egroup}
\let\markbothOrig\markboth
\def\markboth#1#2{\bgroup
  \let\texttt=\textttOrig\markbothOrig{#1}{#2}\egroup}
\let\markrightOrig\markright
\def\markright#1{\bgroup
  \let\texttt=\textttOrig\markrightOrig{#1}\egroup}

\usepackage{graphicx,grffile}
\makeatletter
\def\maxwidth{\ifdim\Gin@nat@width>\linewidth\linewidth\else\Gin@nat@width\fi}
\def\maxheight{\ifdim\Gin@nat@height>\textheight\textheight\else\Gin@nat@height\fi}
\makeatother
\setkeys{Gin}{width=\maxwidth,height=\maxheight,keepaspectratio}
\IfFileExists{parskip.sty}{%
\usepackage{parskip}
}{
\setlength{\parindent}{0pt}
\setlength{\parskip}{6pt plus 2pt minus 1pt}
}
\ifx\paragraph\undefined\else
\let\oldparagraph\paragraph
\renewcommand{\paragraph}[1]{\oldparagraph{#1}\mbox{}}
\fi
\ifx\subparagraph\undefined\else
\let\oldsubparagraph\subparagraph
\renewcommand{\subparagraph}[1]{\oldsubparagraph{#1}\mbox{}}
\fi

\title{starry\_process: Interpretable Gaussian processes for stellar
light curves}

        \author[1, 2]{Rodrigo Luger}
          \author[1]{Daniel Foreman-Mackey}
          \author[3, 4]{Christina Hedges}
    
      \affil[1]{Center for Computational Astrophysics, Flatiron
Institute, New York, NY}
      \affil[2]{Virtual Planetary Laboratory, University of Washington,
Seattle, WA}
      \affil[3]{Bay Area Environmental Research Institute, P.O. Box 25,
Moffett Field, CA 94035, USA}
      \affil[4]{NASA Ames Research Center, Moffett Field, CA}
  \date{\vspace{-7ex}}

\begin{document}
\maketitle

\marginpar{

  \begin{flushleft}
  \sffamily\small

  {\bfseries DOI:} \href{https://doi.org/DOI unavailable}{\color{linky}{DOI unavailable}}

  \vspace{2mm}

  {\bfseries Software}
  \begin{itemize}
    \setlength\itemsep{0em}
    \item \href{N/A}{\color{linky}{Review}} \ExternalLink
    \item \href{NO_REPOSITORY}{\color{linky}{Repository}} \ExternalLink
    \item \href{DOI unavailable}{\color{linky}{Archive}} \ExternalLink
  \end{itemize}

  \vspace{2mm}

  \par\noindent\hrulefill\par

  \vspace{2mm}

  {\bfseries Editor:} \href{https://example.com}{Pending
Editor} \ExternalLink \\
  \vspace{1mm}
    {\bfseries Reviewers:}
  \begin{itemize}
  \setlength\itemsep{0em}
    \item \href{https://github.com/Pending Reviewers}{@Pending
Reviewers}
    \end{itemize}
    \vspace{2mm}

  {\bfseries Submitted:} N/A\\
  {\bfseries Published:} N/A

  \vspace{2mm}
  {\bfseries License}\\
  Authors of papers retain copyright and release the work under a Creative Commons Attribution 4.0 International License (\href{http://creativecommons.org/licenses/by/4.0/}{\color{linky}{CC BY 4.0}}).

  \end{flushleft}
}

\hypertarget{section}{%
\section{}\label{section}}

\hypertarget{summary}{%
\section{Summary}\label{summary}}

\begin{figure}
\centering
\includegraphics{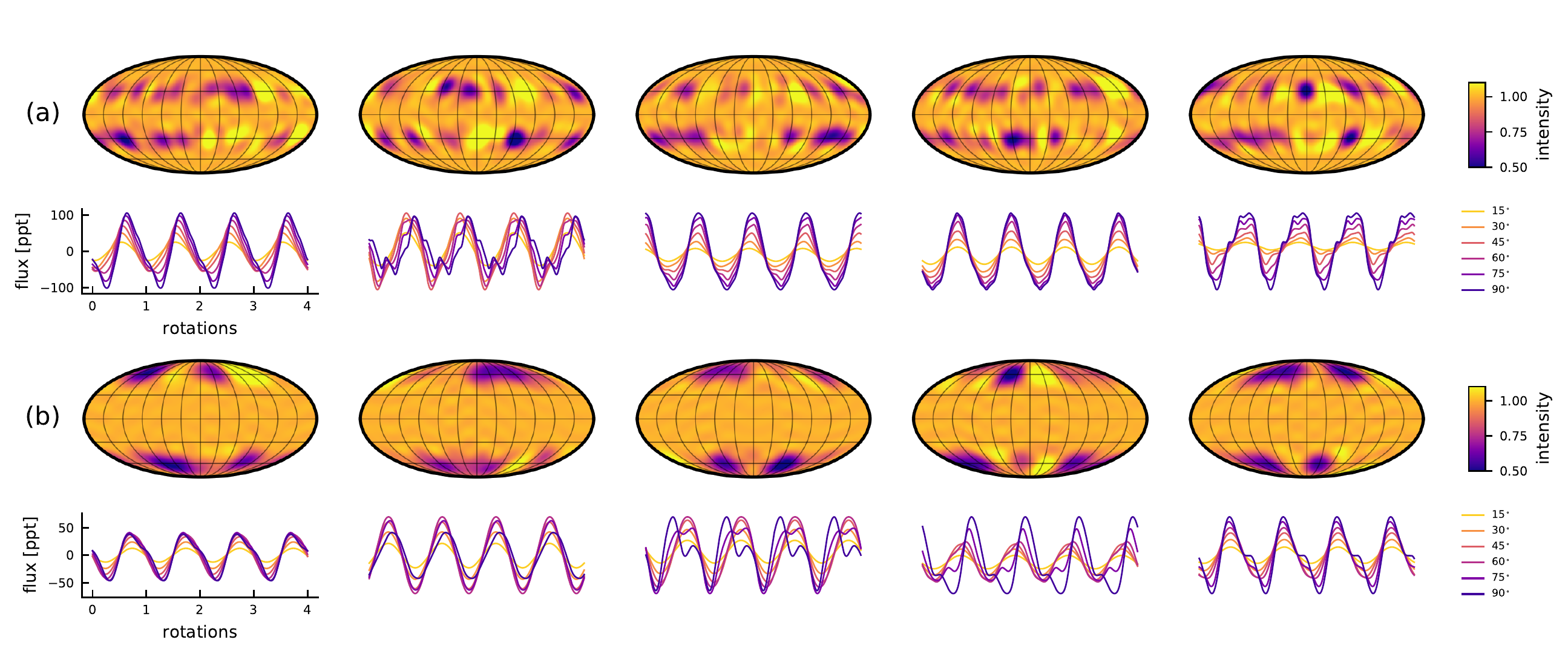}
\caption{Five random samples from our GP (columns) conditioned on two
different hyperparameter vectors \(\pmb{\theta}_\bullet\) (rows). The
samples are shown on the surface of the star in a Mollweide projection
alongside the corresponding light curves viewed at several different
inclinations. (a) Samples from a GP describing a star with small
mid-latitude spots. (b) Samples from a GP describing a star with larger
high-latitude spots. \label{fig:samples}}
\end{figure}

The \texttt{starry\_process} code implements an interpretable Gaussian
process (GP) for modeling variability in stellar light curves. As dark
starspots rotate in and out of view, the total flux received from a
distant star will change over time. Unresolved flux time series
therefore encode information about the spatial structure of features on
the stellar surface. The \texttt{starry\_process} software package
allows one to easily model the flux variability due to starspots,
whether one is interested in understanding the properties of these spots
or marginalizing over the stellar variability when it is treated as a
nuisance signal. The main difference between the GP implemented here and
typical GPs used to model stellar variability is the explicit dependence
of our GP on physical properties of the star, such as its period,
inclination, and limb darkening coefficients, and on properties of the
spots, such as their radius and latitude distributions (see Figure
\ref{fig:samples}). This code is the \texttt{Python} implementation of
the interpretable GP algorithm developed in Luger, Foreman-Mackey, \&
Hedges (2021).

\hypertarget{section-1}{%
\section{}\label{section-1}}

\hypertarget{statement-of-need}{%
\section{Statement of need}\label{statement-of-need}}

Mapping the surfaces of stars using time series measurements is a
fundamental problem in modern time-domain stellar astrophysics. This
inverse problem is ill-posed and computationally intractable, but in the
associated AAS Journals publication submitted in parallel to this paper
(Luger, Foreman-Mackey, \& Hedges, 2021), we derive an interpretable
effective Gaussian Process (GP) model for this problem that enables
robust probabilistic characterization of stellar surfaces using
photometric time series observations. Our model builds on previous work
by Perger et al. (2020) on semi-interpretable Gaussian processes for
stellar timeseries data and by Morris (2020) on approximate inference
for large ensembles of stellar light curves. Implementation of our model
requires the efficient evaluation of a set of special functions and
recursion relations that are not readily available in existing
probabilistic programming frameworks. The \texttt{starry\_process}
package provides the necessary elements to perform this analysis with
existing and forthcoming astronomical datasets.

\hypertarget{implementation}{%
\section{Implementation}\label{implementation}}

We implement our interpretable GP in the user-friendly \texttt{Python}
package \texttt{starry\_process}, which can be installed via
\texttt{pip} or from source on
\href{https://github.com/rodluger/starry_process}{GitHub}. The code is
thoroughly
\href{https://github.com/rodluger/starry_process/tree/master/tests}{unit-tested}
and \href{https://starry_process.readthedocs.io}{well documented}, with
\href{https://starry-process.readthedocs.io/en/latest/examples}{examples}
on how to use the GP in custom inference problems. As discussed in the
associated AAS Journals publication (Luger, Foreman-Mackey, \& Hedges,
2021), users can choose, among other options, whether or not to
marginalize over the stellar inclination and whether or not to model a
normalized process. Users can also choose the spherical harmonic degree
of the expansion, although it is recommended to use
\(l_\mathrm{max} = 15\) (see below). Users may compute the mean vector
and covariance matrix in either the spherical harmonic basis or the flux
basis, or they may sample from it or use it to compute marginal
likelihoods. Arbitrary order limb darkening is implemented following
Agol et al. (2020).

The code was designed to maximize the speed and numerical stability of
the computation. Although the computation of the GP covariance involves
many layers of nested sums over spherical harmonic coefficients, these
may be expressed as high-dimensional tensor products, which can be
evaluated efficiently on modern hardware. Many of the expressions can
also be either pre-computed or computed recursively. To maximize the
speed of the algorithm, the code is implemented in hybrid
\texttt{C++}/\texttt{Python} using the just-in-time compilation
capability of the \texttt{Theano} package (Theano Development Team,
2016). Since all equations derived here have closed form expressions,
these can be autodifferentiated in a straightforward and numerically
stable manner, enabling the computation of backpropagated gradients
within \texttt{Theano}. As such, \texttt{starry\_process} is designed to
work out-of-the box with \texttt{Theano}-based inference tools such as
\texttt{PyMC3} for NUTS/HMC or ADVI sampling (Salvatier et al., 2016).

\begin{figure}
\centering
\includegraphics{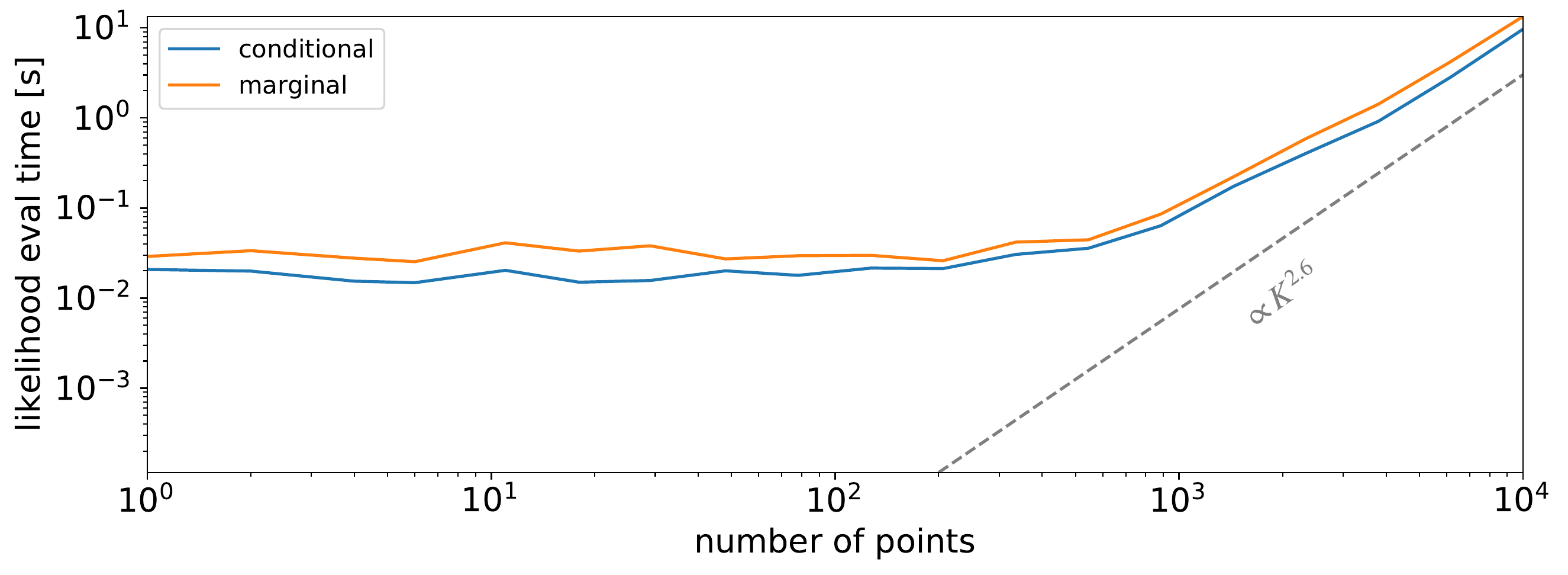}
\caption{Evaluation time in seconds for a single log-likelihood
computation as a function of the number of points \(K\) in each light
curve when conditioning on a value of the inclination (blue) and when
marginalizing over the inclination (orange). At \(l_\mathrm{max} = 15\),
computation of the covariance matrix of the GP takes about 20ms on a
typical laptop. The dashed line shows the asymptotic scaling of the
algorithm, which is due to the Cholesky factorization and solve
operations.\label{fig:speed}}
\end{figure}

Figure \ref{fig:speed} shows the computational scaling of the
\texttt{Python} implementation of the algorithm for the case where we
condition the GP on a specific value of the inclination (blue) and the
case where we marginalize over inclination (orange). Both curves show
the time in seconds to compute the likelihood (averaged over many trials
to obtain a robust estimate) as a function of the number of points \(K\)
in a single light curve. For \(K \lesssim 100\), the computation time is
constant at \(10-30\) ms for both algorithms. This is the approximate
time (on a typical modern laptop) taken to compute the GP covariance
matrix given a set of hyperparameters \(\pmb{\theta}_\bullet\). For
larger values of \(K\), the cost approaches a scaling of \(K^{2.6}\),
which is dominated by the factorization of the covariance matrix and the
solve operation to compute the likelihood. The likelihood marginalized
over inclination is only slightly slower to compute, thanks to the
tricks discussed in Luger, Foreman-Mackey, \& Hedges (2021).

\begin{figure}
\centering
\includegraphics{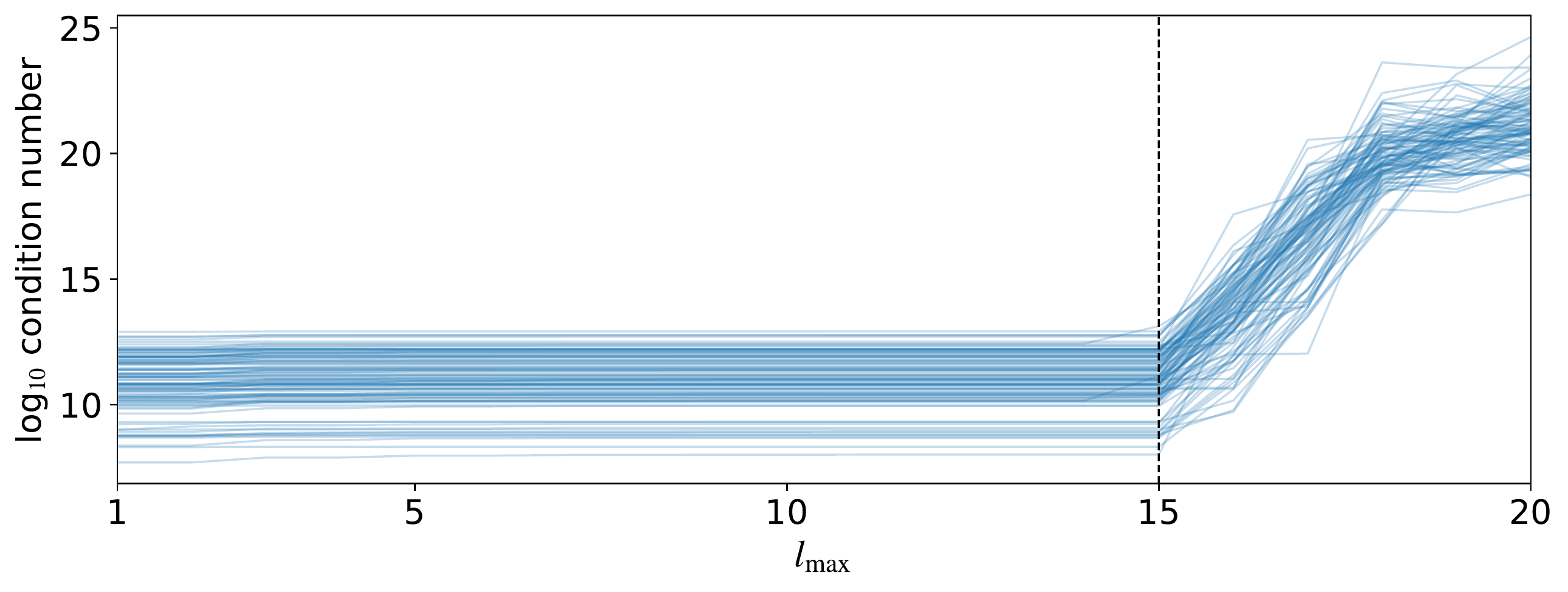}
\caption{Log of the condition number of the covariance in the spherical
harmonic basis as a function of the spherical harmonic degree of the
expansion, \(l_\mathrm{max}\). Different lines correspond to different
values of \(\pmb{\theta}_\bullet\) drawn from a uniform prior (see text
for details). In the majority of the cases, the matrix becomes
ill-conditioned above \(l_\mathrm{max} = 15\).\label{fig:stability}}
\end{figure}

Many modern GP packages (e.g., Ambikasaran et al., 2015; Foreman-Mackey
et al., 2017) have significantly better asymptotic scalings, but these
are usually due to specific structure imposed on the kernel functions,
such as the assumption of stationarity. Our kernel structure is
determined by the physics (or perhaps more accurately, the geometry) of
stellar surfaces, and its nonstationarity is a consequence of the
normalization step in relative photometry (Luger, Foreman-Mackey,
Hedges, \& Hogg, 2021). Moreover, and unlike the typical kernels used
for GP regression, our kernel is a nontrivial function of the
hyperparameters \(\pmb{\theta}_\bullet\), so its computation is
necessarily more expensive. Nevertheless, the fact that our GP may be
used for likelihood evaluation in a small fraction of a second for
typical datasets (\(K \sim 1{,000}\)) makes it extremely useful for
inference.

Our algorithm is also numerically stable over nearly all of the prior
volume up to \(l_\mathrm{max} = 15\). Figure \ref{fig:stability} shows
the log of the condition number of the covariance matrix in the
spherical harmonic basis as a function of the spherical harmonic degree
of the expansion for 100 draws from a uniform prior over the domain of
the hyperparameters. The condition number is nearly constant up to
\(l_\mathrm{max} = 15\) in almost all cases; above this value, the
algorithm suddenly becomes unstable and the covariance is
ill-conditioned. The instability occurs within the computation of the
latitude and longitude moment integrals and is likely due to the large
number of operations involving linear combinations of hypergeometric and
gamma functions. While it may be possible to achieve stability at higher
values of \(l_\mathrm{max}\) via careful reparametrization of some of
those equations, we find that \(l_\mathrm{max} = 15\) is high enough for
most practical purposes.

\hypertarget{references}{%
\section*{References}\label{references}}
\addcontentsline{toc}{section}{References}

\hypertarget{refs}{}
\begin{CSLReferences}{1}{0}
\leavevmode\hypertarget{ref-Agol2020}{}%
Agol, E., Luger, R., \& Foreman-Mackey, D. (2020). {Analytic Planetary
Transit Light Curves and Derivatives for Stars with Polynomial Limb
Darkening}. \emph{159}(3), 123.
\url{https://doi.org/10.3847/1538-3881/ab4fee}

\leavevmode\hypertarget{ref-Ambikasaran2015}{}%
Ambikasaran, S., Foreman-Mackey, D., Greengard, L., Hogg, D. W., \&
O'Neil, M. (2015). {Fast Direct Methods for Gaussian Processes}.
\emph{IEEE Transactions on Pattern Analysis and Machine Intelligence},
\emph{38}, 252. \url{https://doi.org/10.1109/TPAMI.2015.2448083}

\leavevmode\hypertarget{ref-ForemanMackey2017}{}%
Foreman-Mackey, D., Agol, E., Ambikasaran, S., \& Angus, R. (2017).
{Fast and Scalable Gaussian Process Modeling with Applications to
Astronomical Time Series}. \emph{154}(6), 220.
\url{https://doi.org/10.3847/1538-3881/aa9332}

\leavevmode\hypertarget{ref-PaperII}{}%
Luger, R., Foreman-Mackey, D., \& Hedges, C. (2021). {Mapping stellar
surfaces II: An interpretable Gaussian process model for light curves}.
\emph{arXiv e-Prints}, arXiv:2102.01697.
\url{http://arxiv.org/abs/2102.01697}

\leavevmode\hypertarget{ref-PaperI}{}%
Luger, R., Foreman-Mackey, D., Hedges, C., \& Hogg, D. W. (2021).
{Mapping stellar surfaces I: Degeneracies in the rotational light curve
problem}. \emph{arXiv e-Prints}, arXiv:2102.00007.
\url{http://arxiv.org/abs/2102.00007}

\leavevmode\hypertarget{ref-Morris2020b}{}%
Morris, B. (2020). {fleck: Fast approximate light curves for starspot
rotational modulation}. \emph{The Journal of Open Source Software},
\emph{5}(47), 2103. \url{https://doi.org/10.21105/joss.02103}

\leavevmode\hypertarget{ref-Perger2020}{}%
Perger, M., Anglada-Escudé, G., Ribas, I., Rosich, A., Herrero, E., \&
Morales, J. C. (2020). {Auto-correlation functions of astrophysical
processes, and their relation to Gaussian processes; Application to
radial velocities of different starspot configurations}. \emph{arXiv
e-Prints}, arXiv:2012.01862. \url{http://arxiv.org/abs/2012.01862}

\leavevmode\hypertarget{ref-Salvatier2016}{}%
Salvatier, J., Wiecki, T. V., \& Fonnesbeck, C. (2016). {Probabilistic
programming in Python using PyMC3}. \emph{PeerJ Computer Science},
\emph{2}, e55.

\leavevmode\hypertarget{ref-Theano2016}{}%
Theano Development Team. (2016). {Theano: A {Python} framework for fast
computation of mathematical expressions}. \emph{arXiv e-Prints},
\emph{abs/1605.02688}. \url{http://arxiv.org/abs/1605.02688}

\end{CSLReferences}

\end{document}